\documentclass[aps,pre,preprint,showpacs,groupedaddress,superscriptaddress, amsmath,amssymb,nofootinbib]{revtex4-1}
\usepackage[pdftex]{graphicx}

\usepackage{amsmath}
\usepackage{amssymb}
\usepackage{amsfonts}
\usepackage{dcolumn}% Align table columns on decimal point
\usepackage{bm}% bold math
\usepackage{color}

\begin{document}

\title{Role of the single-particle dynamics in the transverse current autocorrelation function of a liquid metal}

\author{Eleonora Guarini}
\affiliation{Dipartimento di Fisica e Astronomia, Universit\`a degli Studi di Firenze, via G. Sansone 1, I-50019 Sesto Fiorentino, Italy}

\author{Ubaldo Bafile}
\affiliation{Consiglio Nazionale delle Ricerche, Istituto di Fisica Applicata ``Nello Carrara'', via Madonna del Piano 10, I-50019 Sesto Fiorentino, Italy}

\author{Daniele Colognesi}
\affiliation{Consiglio Nazionale delle Ricerche, Istituto di Fisica Applicata ``Nello Carrara'', via Madonna del Piano 10, I-50019 Sesto Fiorentino, Italy}

\author{Alessandro Cunsolo}
\affiliation{Department of Physics, University of Wisconsin at Madison, 1150 University Avenue, Madison, WI, 53706, United States}

\author{Alessio De Francesco}
\affiliation{CNR-IOM \& INSIDE@ILL c/o Operative Group in Grenoble (OGG) F-38042 Grenoble, France
and Institut Laue Langevin (ILL), F-38042 Grenoble, France}

\author{Ferdinando Formisano}
\affiliation{CNR-IOM \& INSIDE@ILL c/o Operative Group in Grenoble (OGG) F-38042 Grenoble, France
and Institut Laue Langevin (ILL), F-38042 Grenoble, France}

\author{Wouter Montfrooij}
\affiliation{Department of Physics and Astronomy, University of Missouri, Columbia, MO, 65211, United States}

\author{Martin Neumann}
\affiliation{Fakult\"{a}t f\"{u}r Physik der Universit\"{a}t Wien, Kolingasse 14-16, A-1090 Wien, Austria}

\author{Fabrizio Barocchi}
\affiliation{Dipartimento di Fisica e Astronomia, Universit\`a degli Studi di Firenze, via G. Sansone 1, I-50019 Sesto Fiorentino, Italy}

\begin{abstract}

A very recent simulation study of the transverse current autocorrelation of the Lennard-Jones fluid revealed, as expected, that this function can be perfectly described within the exponential expansion theory. However, above a certain wavevector $Q$, not only transverse collective excitations are found to propagate in the fluid, but a second oscillatory component of unclear origin (thereby called X) must be considered to properly account for the time behavior of the correlation. Here we present an extended investigation of the transverse current autocorrelation of liquid Au as obtained by ab initio molecular dynamics in the very wide range 5.7 nm$^{-1}$ $\le Q \le$ 32.8 nm$^{-1}$ in order to follow the behavior of the X component, if present, also at large $Q$ values. By combining the study of the transverse current autocorrelation with the analogous analysis of its self part, we show that the second oscillatory component originates from the longitudinal dynamics and appears in the same form as a collective excitation is represented in the single-particle behavior. Therefore, the signature of the longitudinal processes (sound waves) in the transverse current autocorrelation is not due to often conjectured couplings of longitudinal and trasverse modes, but descends from the self part of the function, which contains the traces of all processes acting in the fluid as the density of states, that is the spectrum of the velocity autocorrelation function, does.     

\end{abstract}

%\date{\today}

\maketitle

\section{Introduction}
\label{sect: intro}

Liquid metals have always been considered as reference systems for investigations of the microscopic dynamics of liquids, due to their monatomic nature and to the typically intense features of the spectrum of the van Hove density-density correlation function, i.e., the dynamic structure factor $S(Q,\omega)$ \cite{march, balucani, montfrooij}. Early studies mainly focused on the characterization of longitudinal collective excitations (often referred to as ``sound waves'') propagating in these fluids \cite{montfrooij, scopigno_review,guarini2013}. More recently, experimental and simulation inquiries of the dynamics of liquid metals mostly addressed the behavior of transverse excitations (``shear waves''), certainly present in dense fluids at sufficiently small wavelengths, i.e., above a threshold wavevector $Q$ value which, as observed from ab initio molecular dynamics (AIMD) simulations of these systems, can be as low as a few inverse nanometers \cite{marques2015,delrio2016,delrio2017,delrio2017a, delriozinco}. Indeed, simulations provide, at present, the only possibility to determine the most crucial functions for studies of the transverse dynamics: the transverse current autocorrelation function (TCAF) $C_{\rm T}(Q,t)$ and the velocity autocorrelation function (VAF) here indicated as $Z(t)$ \cite{balucani}. The former gives direct information on transverse excitations with varying $Q$, the latter is instead a function of time only to which both longitudinal and transverse collective processes contribute by involving the motion, and thus affecting the velocity, of each single particle. In particular, the spectrum of the VAF $\tilde{Z}(\omega)$ can be considered to represent, for a liquid, the equivalent of the phonon density of states (DoS) of a solid, thereby revealing, in an indirect way, all the excitations sustained by the fluid, as clearly shown in the literature both for a model Lennard-Jones (LJ) dense fluid \cite{bellissima2017} and for liquid metals \cite{guarini2017,guarini2020}.

Very recently, we showed that by using the exponential expansion theory (EET) of correlation functions \cite{barocchi2012, barocchi2013, barocchi2014} a remarkably accurate account of the TCAF and of its spectrum $\tilde{C}_{\rm T}(Q,\omega)$ of a LJ fluid can be obtained \cite{guarini2023}. In that work, the studied thermodynamic state and $Q$ range allowed, in particular, to accurately locate,  thanks to the EET decomposition, the wavevector at which shear waves start to propagate and to relate this process to a damped harmonic oscillator smoothly undergoing a transition from an over- to an underdamped state. Unexpectedly, at higher $Q$ values, we also found that an additional oscillator, performing an equally continuous change from over- to underdamped conditions, was required to properly describe $C_{\rm T}(Q,t)$. The oscillation frequency of this second underdamped component (labeled as X in Ref.\ \cite{guarini2023}) was observed to grow steeply with $Q$, rapidly overtaking the one of transverse waves. However, the available data did not provide enough elements to confidently draw conclusions about the nature and physical meaning of the X contribution to $C_{\rm T}(Q,t)$. Nonetheless, its frequency $\omega_{\rm X}$ appeared to grow with $Q$ towards the value of the maximum of the dispersion curve of the longitudinal acoustic excitation obtained \cite{bellissima2017} for the same thermodynamic state of the LJ fluid (see, in particular, the red curve in Fig.\ 9 of Ref.\ \cite{bellissima2017}). This fact might lead to hypothesize the X ``propagating collective excitation'' as a possible fingerprint in $C_{\rm T}(Q,t)$ of the longitudinal dynamics (the reason of the quotation marks will be elucidated in the remainder of the paper). 
 
In order to get insight about the origin of this second phenomenon, we turn here to the analysis of $C_{\rm T}(Q,t)$ of liquid Au as obtained from the simulated atomic configurations already used to calculate both the total dynamic structure factor $S(Q,\omega)$ \cite{guarini2013} and its single-particle (self) part $S_{\rm self}(Q,\omega)$ \cite{guarini2017}. The present investigation is extended to rather high $Q$ values, thus allowing us to follow the frequency and damping of the exponential modes well beyond
$Q_{\rm p}$, i.e., the position of the main peak of the static structure factor ($Q_{\rm p}$ = 26 nm$^{-1}$ for Au). In this way, we could check the presence of the X component in $C_{\rm T}(Q,t)$ of a system greatly differing from the LJ fluid, and follow its evolution in a wide $Q$ range. The choice of Au was suggested not only by the availability of reliable and well-tested \cite{guarini2013} AIMD simulations, but also, as mentioned, by the enhanced dynamical features typically characterizing correlation functions of liquid metals with respect to other simple fluids. In fact, the more marked dynamical behavior helps, in general, an easier understanding of the various properties. Moreover, the monatomic nature ensures the absence of optic-like modes (e.g., with an intramolecular character) that might make the interpretation more complex, and allows us to focus solely on acoustic excitations.

The present work will bring quite a convincing proof of the longitudinal origin of the X mode of $C_{\rm T}(Q,t)$, which is clearly detected in the case of simulated Au too, as in the model LJ case. However, it is important to anticipate that such an origin must be intended in the very special sense we are going to clarify. Indeed, traces of the longitudinal dynamics in a transverse correlation should not be interpreted as some evidence of a mixing/coupling of the longitudinal and transverse excitations. This mixing has often been conjectured (see e.g., Ref.\ \cite{brazhkin} and references therein), though never quantitatively demonstrated or theoretically derived, in order to attempt an explanation of the fact that signs of transverse waves in $S(Q,\omega)$ and longitudinal waves in the spectrum of the TCAF, $\tilde{C}_{\rm T}(Q,\omega)$, have been observed in some simulated liquid metals or, mainly, hydrogen bonded liquids \cite{sampoli97}. We propose a different interpretation, independent of the coupling concept, of the reciprocal signatures of the main collective processes of fluids in specialized correlation functions, that is, in functions most appropriate to characterize either the longitudinal dynamics ($S(Q,\omega)$) or the transverse one ($\tilde{C}_{\rm T}(Q,\omega)$).

\section{Basic definitions and preliminary observations}
\label{prelim}

The current ${\bf j}$ is defined as ${\bf j}({\bf Q},t)=\sum_\alpha{\bf v}_\alpha(t) e^{i{\bf Q}\cdot{\bf R}_\alpha(t)}$, with ${\bf R}_\alpha(t)$ and ${\bf v}_\alpha(t)$ indicating the position and velocity of the $\alpha$-th particle. The current can be separated in two contributions, ${\bf j}_{\rm L}$ and ${\bf j}_{\rm T}$, where the longitudinal component (parallel to $ {\bf Q}$) is given by \cite{balucani}
 
\begin{equation}
\label{jL}
{\bf j}_{\rm L}({\bf Q},t)=({\bf j}({\bf Q},t)\cdot {\bf Q}){\bf Q}/Q^2,
\end{equation}

\noindent and the transverse one is simply obtained by taking the difference ${\bf j}_{\rm T}({\bf Q},t)={\bf j}({\bf Q},t)-{\bf j}_{\rm L}({\bf Q},t)$. Following the notation of Ref.\ \cite{balucani} for classical systems, the longitudinal current autocorrelation function (LCAF) is then

\begin{equation}
\label{CL}
C_{\rm L}(Q,t)=\frac{1}{N}\langle{\bf j}_{\rm L}^*({\bf Q},0) \cdot {\bf j}_{\rm L}({\bf Q},t) \rangle=-\frac{1}{Q^2} \frac{d^2 F(Q,t)}{dt^2},
\end{equation}

\noindent where $F(Q,t)=\frac{1}{N} \sum_{\alpha,\beta}\langle e^{-i {\bf Q} \cdot {\bf R}_\alpha(0)}e^{i {\bf Q} \cdot {\bf R}_\beta(t)}\rangle$ is the intermediate scattering function. The TCAF is instead given by

\begin{equation}
\label{CT}
C_{\rm T}(Q,t)=\frac{1}{2N}\langle{\bf j}_{\rm T}^*({\bf Q},0) \cdot {\bf j}_{\rm T}({\bf Q},t) \rangle.
\end{equation}

\noindent In the above equations $N$ is the number of atoms and $\langle \dots \rangle$ denotes, as usual, the ensemble average.

By assuming ${\bf Q}$ parallel to the $z$-axis, one can also write the LCAF as \cite{balucani}:

\begin{equation}
\begin{split}
\label{correntelong}
C_{\rm L}(Q,t)=\frac{1}{N}\langle j^*_{z}({\bf Q},0) j_{z}({\bf Q},t) \rangle =\frac{1}{N}\sum_{\alpha,\beta \ne \alpha}\langle v_{z,\alpha}(0) e^{-iQ R_{z,\alpha}(0)} v_{z,\beta}(t) e^{iQ R_{z,\beta}(t)}\rangle +\\
\frac{1}{N}\sum_{\alpha} \langle v_{z,\alpha}(0) e^{-iQ R_{z,\alpha}(0)}v_{z,\alpha}(t) e^{iQ R_{z,\alpha}(t)}\rangle =C_{\rm L,dist}(Q,t)+C_{\rm L,self}(Q,t),
\end{split}
\end{equation}

\noindent where we introduced the self and distinct components of the function. The same separation can be operated for the TCAF:

\begin{equation}
\begin{split}
\label{correntetrasv}
C_{\rm T}(Q,t)=\frac{1}{N}\langle j^*_{x}({\bf Q},0) j_{x}({\bf Q},t)\rangle =\frac{1}{N}\sum_{\alpha,\beta \ne \alpha}\langle v_{x,\alpha}(0) e^{-iQ R_{z,\alpha}(0)}v_{x,\beta}(t) e^{iQ R_{z,\beta}(t)} \rangle +\\
\frac{1}{N}\sum_{\alpha} \langle v_{x,\alpha}(0) e^{-iQ R_{z,\alpha}(0)}v_{x,\alpha}(t) e^{iQ R_{z,\alpha}(t)} \rangle =C_{\rm T,dist}(Q,t)+C_{\rm T,self}(Q,t).
\end{split}
\end{equation}

\noindent Of course, due to the isotropy of the fluid, one also has $C_{\rm T}(Q,t)=\frac{1}{N}\langle j^*_{x}({\bf Q},0) j_{x}({\bf Q},t) \rangle=\frac{1}{N}\langle j^*_{y}({\bf Q},0) j_{y}({\bf Q},t)\rangle$.

As far as the VAF is concerned, it is given by

\begin{equation}
\label{VAF} 
Z(t)=\frac{1}{N}\sum_\alpha\langle {\bf v}_\alpha(0) \cdot {\bf v}_\alpha(t) \rangle.
\end{equation}
 
\noindent Therefore, comparison of Eqs.\ (\ref{correntelong}), (\ref{correntetrasv}), and (\ref{VAF}) readily reveals that, in the $Q \to 0$ limit, the following equalities hold

\begin{equation}
\label{leself} 
C_{\rm L,self}(Q \to 0,t)=C_{\rm T,self}(Q \to 0,t)=\frac{1}{3}Z(t).
\end{equation}

\noindent As we will show, the self part of the TCAF has a very weak dependence on $Q$. As a consequence, $C_{\rm T}(Q,t)$ contains the distinct part and a contribution which is essentially very similar to the VAF, even at nonzero $Q$ values. Correspondingly, $\tilde{C}_{\rm T}(Q,\omega)$ has a spectral component that brings the same information of the DoS of the fluid. 

It is important to recall that an EET representation of the VAF allows one to distinguish the longitudinal and transverse contributions to the DoS \cite{guarini2017}. However, the characteristic frequencies derived from the analysis of the VAF do not correspond, strictly speaking, to those of ``propagating collective excitations'' in the implied typical sense, linked also to the concept of dispersion. In fact, $\tilde{Z}(\omega)$ is independent of $Q$ and has peaks or shoulders at frequencies where the branches of the dispersion relation have a horizontal tangent, in agreement with its physical meaning of being a density of states. For strongly dispersive collective excitations, like the longitudinal ones, the DoS of a liquid metal displays a broad shoulder around a frequency corresponding to the maximum of the longitudinal dispersion curve $\omega_{\rm s}(Q)$ (the subscript s meaning ``sound'') obtained from the analysis of $S(Q,\omega)$ \cite{guarini2013, guarini2017, guarini2020}, as can be appreciated in Fig.\ \ref{Fig1} for the case of Au. In this sense, the presence of features in the DoS in some frequency bands actually tells us that longitudinal and transverse branches are present in the dispersion relation and where their average is located. Ultimately, the DoS witnesses that both sound and shear waves exist in the fluid.
 
Details of the simulations were given in Ref.\ \cite{guarini2013}. Here it is useful to recall that the simulation was performed with 200 atoms in a cubic box with 1.557 nm edge length, so to give the number density $n$= 53 nm$^{-3}$ of Au slightly above the melting temperature ($T_{\rm m}$ = 1337 K). The above edge length allows for a minimum $Q$ value of 4.0 nm$^{-1}$. From the atomic configurations we calculated both $C_{\rm T,L}(Q,t)$ and $C_{\rm T,L,self}(Q,t)$. The limited variation of $C_{\rm T,self}(Q,t)$ with increasing wavevector is shown in Fig.\ \ref{Fig2}, where we display the self parts of the LCAF and TCAF at two quite different $Q$ values like $4.0$ and $25.5$ nm$^{-1}$, along with $Z(t)/3$. At our minimum $Q$, the self parts of the current correlations are still indistinguishable from $Z(t)/3$ (see Eq.\ (\ref{leself})). By contrast, at the higher $Q$ we observe that $C_{\rm T,self}(Q,t)$ continues to be quite close to $Z(t)/3$, while departures are more evident in the case of $C_{\rm L,self}(Q,t)$.

\section{Analysis of $C_{\rm T}(Q,t)$}
\label{analysisCT}

As mentioned, the simulated $C_{\rm T}(Q,t)$ data were analyzed by means of the EET \cite{barocchi2012, barocchi2013, barocchi2014}, which allows for very good descriptions of various correlation functions and spectra of interest in studies of the self \cite{bellissima2017, guarini2017} and collective dynamics \cite{guarini2020, guarini2021}. The theory predicts that any autocorrelation function can be expressed as a series of exponential terms (called modes). Thus, we write, at each $Q$ value 

\begin{equation}\label{eq:EET}
C_{\rm T}(Q,t)=C_{\rm T}(Q,0)\sum_{j=1}^{\infty}I_j\exp(z_j|t|),
\end{equation}

\noindent where both $I_j$ and $z_j$ can either be real or complex, with ${\rm Re}\,z_j<0$, since it represents the damping coefficient either of relaxation processes or of oscillatory components of $C_{\rm T}(Q,t)$. In the correlation, pure exponential decays are accounted for in the series by what will be referred to as ``real modes", i.e., having both $I_j$ and $z_j$ real. On the other hand, damped oscillatory components of the correlation are represented in the series by what we will designate as ``complex (conjugate) pairs", i.e. by $I_j\exp(z_j t)+I_j^*\exp(z_j^* t)$, with both $I_j$ and $z_j$ complex. In Eq.\ (\ref{eq:EET}), $I_j$ and $z_j$ depend on $Q$, although we omitted this dependence in the above formula.

Details on the application of the EET can be found in Refs.\ \cite{bellissima2017, guarini2017, guarini2020}. The analysis consists in performing a fitting procedure aimed at determining the parameters $z_j$ and $I_j$ of a small number $p$ of modes to which the sum in Eq.\ (\ref{eq:EET}) effectively reduces. Here we only note that $p-1$ constraints have been imposed to the amplitudes $I_j$ in order to enforce the correct short time behavior of the fitted $C_{\rm T}(Q,t)$ \cite{guarini2020}. Since the resulting number of modes turned out to be $p$=4 at all investigated wavevectors, the constraints were $\sum_{j=1}^{p}I_j=1$, which follows directly from Eq.\ (\ref{eq:EET}) at $t=0$, $\sum_{j=1}^{p} I_j z_j=0$, and $\sum_{j=1}^{p} I_j z_j^3=0$, ensuring finite values of the second and fourth spectral moments of $\tilde{C}_{\rm T}(Q,\omega)$.

In detail, we perfomed successful fits at each available $Q$ value in the range 5.7 nm$^{-1} < $ $Q <$ 32.8 nm$^{-1}$,
while, as often happens at the lowest $Q$ value admitted by the simulation box size, the TCAF at 4.0 nm$^{-1}$ turned out to be affected by the boundary conditions and difficult to fit properly.
At the first two $Q$ values of the above range, models containing two real modes and one (low frequency) complex pair were found to provide a very good description of $C_{\rm T}(Q,t)$ in its entire time range, indicating that shear waves have already set in at the wavevectors probed by the simulations. Conversely, at $Q > 7.0$ nm$^{-1}$ an appropriate account of the data could only be obtained by considering no real modes and two complex pairs, meaning that, like in the LJ case, a second (underdamped) oscillatory component (labeled as X, for consistency with Ref.\ \cite{guarini2023}) contributes to $C_{\rm T}(Q,t)$, together with the transverse one. Note that in presence of underdamped oscillatory components, we will use the symbols $\omega_j$, $\Gamma_j$ in place of ${\rm Im}z_j$,$-{\rm Re}z_j$, respectively.

Before showing the $Q$ dependence of the frequency and damping of these pairs of modes, we provide in Fig.\ \ref{Fig3} an example of the quality of the fit to $C_{\rm T}(Q,t)/C_{\rm T}(Q,0)$ at an intermediate $Q$ value. 

Figure \ref{Fig4} shows the $Q$ dependence of the effective frequencies $\omega_{\rm T,X}$, dampings $\Gamma_{\rm T,X}$, and undamped frequencies $\Omega_{\rm T,X}=\sqrt{\omega_{\rm T,X}^2+\Gamma_{\rm T,X}^2}$ of the two contributions to $C_{\rm T}(Q,t)$. Despite the transverse dispersion curve $\omega_{\rm T}(Q)$ in Fig.\ \ref{Fig4}(a) displays a noisy behavior, the trend of $\Omega_{\rm T}(Q)$ in Fig.\ \ref{Fig4}(b) is more regular and actually resembles that observed in liquid Ag \cite{delrio2016,guarini2020}. More comments are worth as far as the X pair is concerned. First of all, we note that the initial $Q$ dependence of $\Omega_{\rm X}$ and $\Gamma_{\rm X}$ is very much the same as the one found in the underdamped state of the LJ case, namely a nearly flat behavior for the former and a nearly linear decrease of the latter. This observation suggests that such trends are general, independently of the specific nature of the fluid. It can also be noted that, at $Q \ge$ 20 nm$^{-1}$, $\Gamma_{\rm X}$ attains the same almost constant value of $\Gamma_{\rm T}$.  

The wide $Q$ range considered in the present paper allows to establish that, after a steep growth, $\omega_{\rm X}(Q)$ reaches the value of the maximum (30 rad ps$^{-1}$ for Au) of the longitudinal dispersion curve $\omega_{\rm s}(Q)$ (see Fig.\ \ref{Fig1}(a)). This behavior was only guessed in the LJ case where the analyzed $Q$ range did not extend to values large enough for a direct observation of its possible limit behavior. Interestingly, here we are able to see that such a frequency value (attained by the X component at $Q \approx$ 18 nm$^{-1}$), which is also the frequency related to the longitudinal processes in the VAF (see Fig.\ \ref{Fig1}(b)), does not change anymore with increasing $Q$. To further check this constant trend we performed a fit to $C_{\rm T}(Q,t)$ also at a value as high as $Q=$ 38.1 nm$^{-1}$, finding (see Fig.\ \ref{Fig4}) for both damping and frequency a behavior similar to that of the preceding $Q$ values. Thus, above 18 nm$^{-1}$, the behavior of $\omega_{\rm X}$ does not correspond to the dispersion of a propagation, as confirmed by what follows.

In Sec.\ \ref{prelim}, we preliminarly noted that the relation of Eq.\ (\ref{leself}), exact at $Q \to 0$, continues to approximately hold for $C_{\rm T, self}(Q,t)$ also at higher wavevectors (see Fig.\ \ref{Fig2}). On the other hand, we now find that $C_{\rm T}(Q,t)$ contains an oscillatory component which, irrespective of the $Q$ value above 18 nm$^{-1}$, seems to be equal to the longitudinal complex pair of the VAF. These observations lead to interpret the X contribution to $C_{\rm T}(Q,t)$ not only as longitudinal in nature, with the same meaning this has for the VAF, but also as representing quite a strong fingerprint in $C_{\rm T}(Q,t)$ of its own self part, thus, ultimately, of the VAF. To quantitatively verify this hypothesis, we found it crucial to perform the EET analysis of $C_{\rm T, self}(Q,t)$ described in the next section.

\section{Analysis of $C_{\rm T,self}(Q,t)$ and discussion of the results}
\label{analysisCTself}

For monatomic fluids, the self part of a correlation function is also a correlation function by itself, characterized by a positive spectrum. The EET can then be applied also to self correlation functions, as already done in Ref.\ \cite{guarini2017}. We thus modeled $C_{\rm T,self}(Q,t)$ according to

\begin{equation}\label{eq:EETself}
C_{\rm T,self}(Q,t)=C_{\rm T,self}(Q,0)\sum_{j=1}^{\infty}I_{j,\rm self}\exp(z_{j,\rm self}|t|),
\end{equation}

\noindent and performed fits in the same $Q$ range investigated for $C_{\rm T}(Q,t)$. Given the close resemblance of $C_{\rm T,self}(Q,t)$ with the VAF also at nonzero wavevector values, the same model adopted in Ref.\ \cite{guarini2017}, foreseeing two complex pairs plus one real mode, was used. As expected, the model proved to be very accurate at all wavevectors. Figure \ref{Fig5} shows its performance at an example $Q$ value, where the higher frequency component is labeled as 2. We will indicate the parameters of the high-frequency mode of $C_{\rm T,self}(Q,t)$ with the symbols $\omega_2$ and $\Gamma_2$ ($\Omega_2=\sqrt{\omega_2^2+\Gamma_2^2}$). Accordingly, for the low-frequency complex pair we use $\omega_1$ and $\Gamma_1$ ($\Omega_1=\sqrt{\omega_1^2+\Gamma_1^2}$).

In Fig.\ \ref{Fig6} the fit results are compared with those of Fig.\ \ref{Fig4}. Very smooth trends of the parameters are observed, with a net superposition \cite{nota}, at intermediate and high $Q$ values, of $\omega_{\rm X}(Q)$ and $\Gamma_{\rm X}(Q)$ of $C_{\rm T}(Q,t)$ with $\omega_2(Q)$ and $\Gamma_2(Q)$ of $C_{\rm T,self}(Q,t)$. Conversely, the transverse dispersion curve $\omega_{\rm T}(Q)$, and even more the undamped frequency $\Omega_{\rm T}(Q)$, expectedly do not coincide with $\omega_1(Q)$ and $\Gamma_1(Q)$ of the self correlation function, except at wavevectors exceeding approximately 25 nm$^{-1}$. 

The two pairs of modes of $C_{\rm T}(Q,t)$ appear then to have profoundly diverse natures, not only because of the different processes in the fluid they are related to (shear and sound waves), but also because the lower frequency complex pair (the transverse one) embodies, at low and intermediate wavevectors, the genuine collective excitation that the correlation function is most appropriate to reveal, while the other is substantially related, at almost all $Q$ values, to the way in which the existence of longitudinal modes is witnessed by a single-particle property. 

These considerations are partly supported by the fact that the distinct part of the correlation plays a role on the transverse modes in the greatest part of the $Q$ range, giving rise to the weak but visible dispersion of the T branch of $C_{\rm T}(Q,t)$. By contrast, the X modes are sensitive to the distinct component only in the first part of the $Q$ range, otherwise their frequency would differ, also above 18 nm$^{-1}$, from what found by fits to $C_{\rm T,self}(Q,t)$. On the other hand, observing that the distinct dynamics mostly affects the transverse excitations means that are exactly {\it these modes} of $C_{\rm T}(Q,t)$ that bring the major information about relative motions of different particles, in agreement with the genuinely collective, propagating, and dispersive character we previously attributed to the T component of the TCAF.

Nonetheless, such a character is eventually lost also for the T mode of the correlation above 25 nm$^{-1}$, where its frequency starts coinciding on average, and within the scattering of the points, with $\omega_1(Q)$. Therefore, this wavevector value marks the limit (in Au) above which nothing can be viewed as probing a strictly ``propagating collective excitation''. Consequently, only the frequencies of $C_{\rm T,self}(Q,t)$, or equivalently of the VAF, can rightly be found also from the total correlation. {\it A posteriori}, one realizes that such a collective to single-particle transition in the character of the T component occurs at a $Q$ value corresponding to a distance $(2 \pi/Q) \simeq$ 0.25 nm, which, at the density of liquid gold, is very close to the average interparticle distance, so that the probed dynamics is essentially that of one atom. Accordingly, going to smaller (and no longer significant at a ``collective level'') length scales cannot actually bring new information, besides that already contained in the VAF. 

For completeness, the comparison of $C_{\rm T,self}(Q,t)$ and $C_{\rm T}(Q,t)$ at 38.1 nm$^{-1}$ is  reported in Fig.\ \ref{Fig7}. The very slight difference between the two curves at times longer than 0.1 ps actually does not entail a change of the frequency (see Fig.\ \ref{Fig6}(b)) as determined by the fits to the two functions, but only a small, likely not significant, difference in the damping (see Fig.\ \ref{Fig6}(c)). In this respect, another important remark is suggested by Fig.\ \ref{Fig8}, where the damping derived from the analysis of $S(Q,\omega)$ (here again indicated as $z_{\rm s}(Q)$ for consistency with Ref.\ \cite{guarini2013}) is superimposed to the results already given in Fig.\ \ref{Fig6}(c). Interestingly, all dampings tend to overlap beyond the well known ``propagation gap'' of $\omega_{\rm s}(Q)$, typically occurring around $Q_{\rm p}$ \cite{guarini2013}. This trend corroborates our overall picture since, at high enough $Q$ values, only a single damping mechanism seems to be detected, whatever dynamical process is investigated through whatever autocorrelation function. Apparently, this is a further proof that, above $Q_{\rm p}$, only the single-particle dynamics is essentially probed.

\section{Synopsis and final remarks}
\label{sect: concl}

This paper was aimed at providing, on a quantitative basis, an interpretation of an unexpected dynamical feature of the TCAF of fluids we recently found in the case of a LJ system: namely, the existence, along with transverse collective excitations, of another oscillatory component of unclear origin (thereby designated as X) in the correlation. The availability of reliable AIMD simulations for liquid Au allowed us to address the case of a ``real'' dense fluid and, at the same time, to span a rather wide $Q$ range, where transverse modes have already set in and, if present, the X modes could be followed appropriately in their evolution with $Q$.

Our analysis, based on the EET of correlation functions, proved again to be very successful as in many other cases already reported, and confirmed the presence of the X modes also in $C_{\rm T}(Q,t)$ of Au, up to the rather high wavevectors of this investigation. The observation of the same phenomenon in so different fluids constitutes an interesting result {\it per se}.

An important hint concerning which work was further required for a better understanding of the nature of the X modes of $C_{\rm T}(Q,t)$ was the observation of their almost constant frequency above a certain $Q$. Moreover, the frequency value was exactly one of those found in previous works on Au both from the analysis of the VAF and from the maxima of the longitudinal dispersion curve derived from $S(Q,\omega)$. 

\begin{figure*}
\resizebox{0.95\textwidth}{!}
{\includegraphics[trim=2cm 10cm 2cm 0cm]{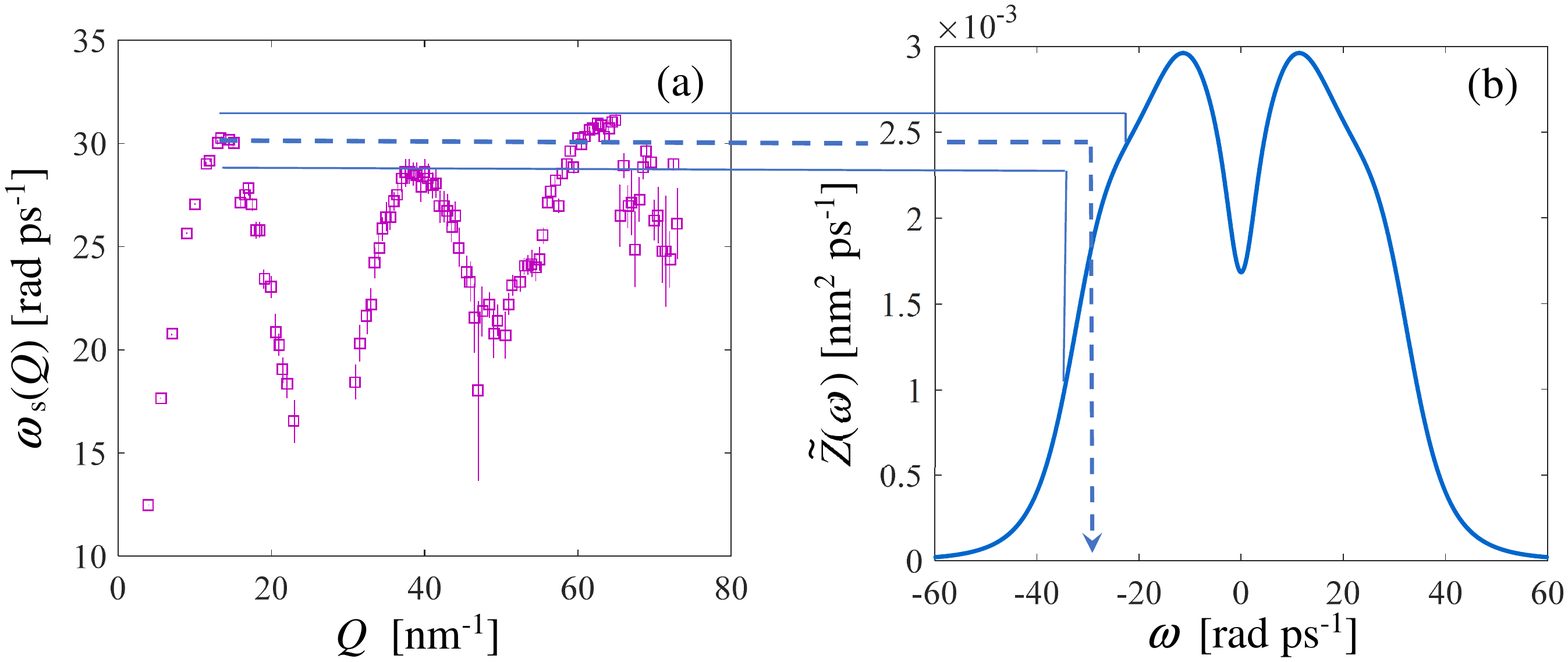}}
\caption{(a) Longitudinal dispersion curve $\omega_{\rm s}(Q)$ of liquid Au as obtained from the analysis of the AIMD $S(Q,\omega)$ \cite{guarini2013}. (b) Spectrum of the VAF (DoS) of liquid Au \cite{guarini2017}. The figure shows that the DoS displays a broad shoulder, centred around 30 ps$^{-1}$ (dashed arrow), in correspondence of the frequency band (rendered by thin solid lines) approximately containing the maxima of $\omega_{\rm s}(Q)$.}
\label{Fig1} 
\end{figure*}

\begin{figure*}
\resizebox{0.95\textwidth}{!}
{\includegraphics[trim=2cm 10cm 2cm 0cm]{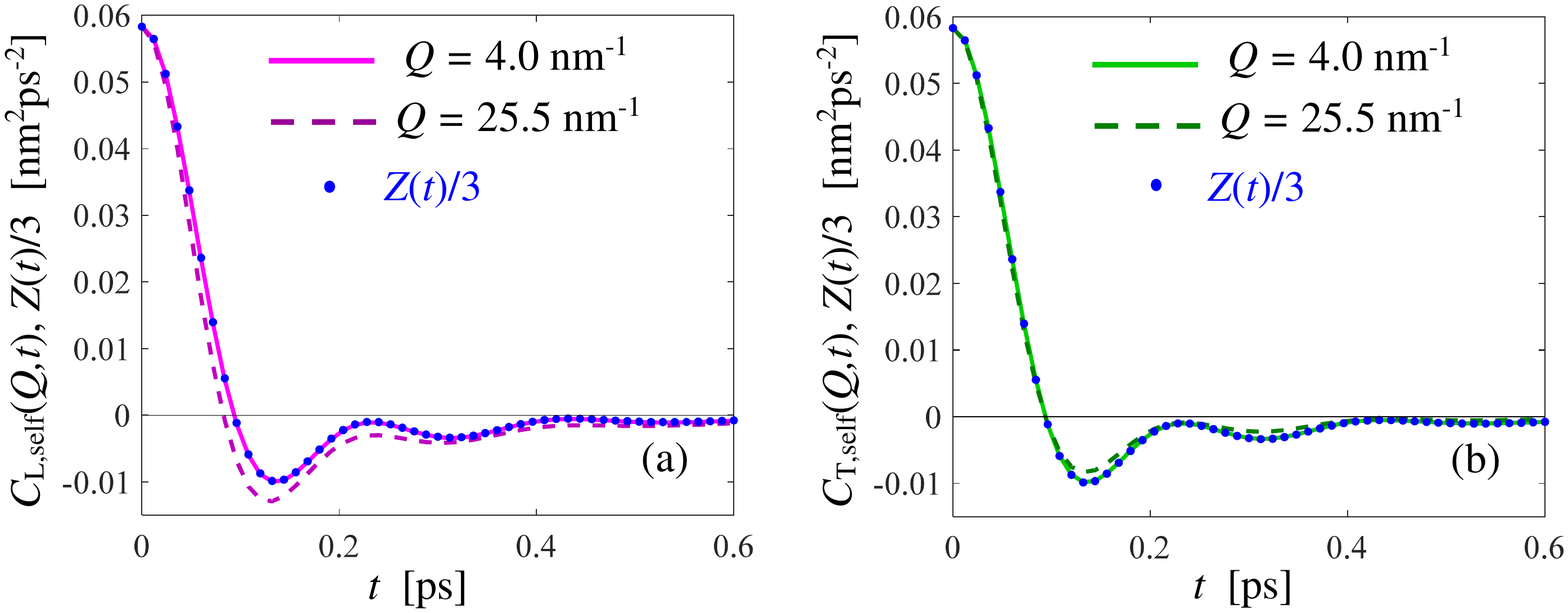}}
\caption{Time dependence of the self parts of (a) the longitudinal and (b) transverse current autocorrelations, compared with $Z(t)/3$ (blue dots). Two $Q$ values are shown for the self current correlations: 4.0 nm$^{-1}$ (solid) and 25.5 nm$^{-1}$ (dashed).}
\label{Fig2} 
\end{figure*}

\begin{figure*}
\resizebox{0.95\textwidth}{!}
{\includegraphics[trim=2cm 10cm 2cm 0cm]{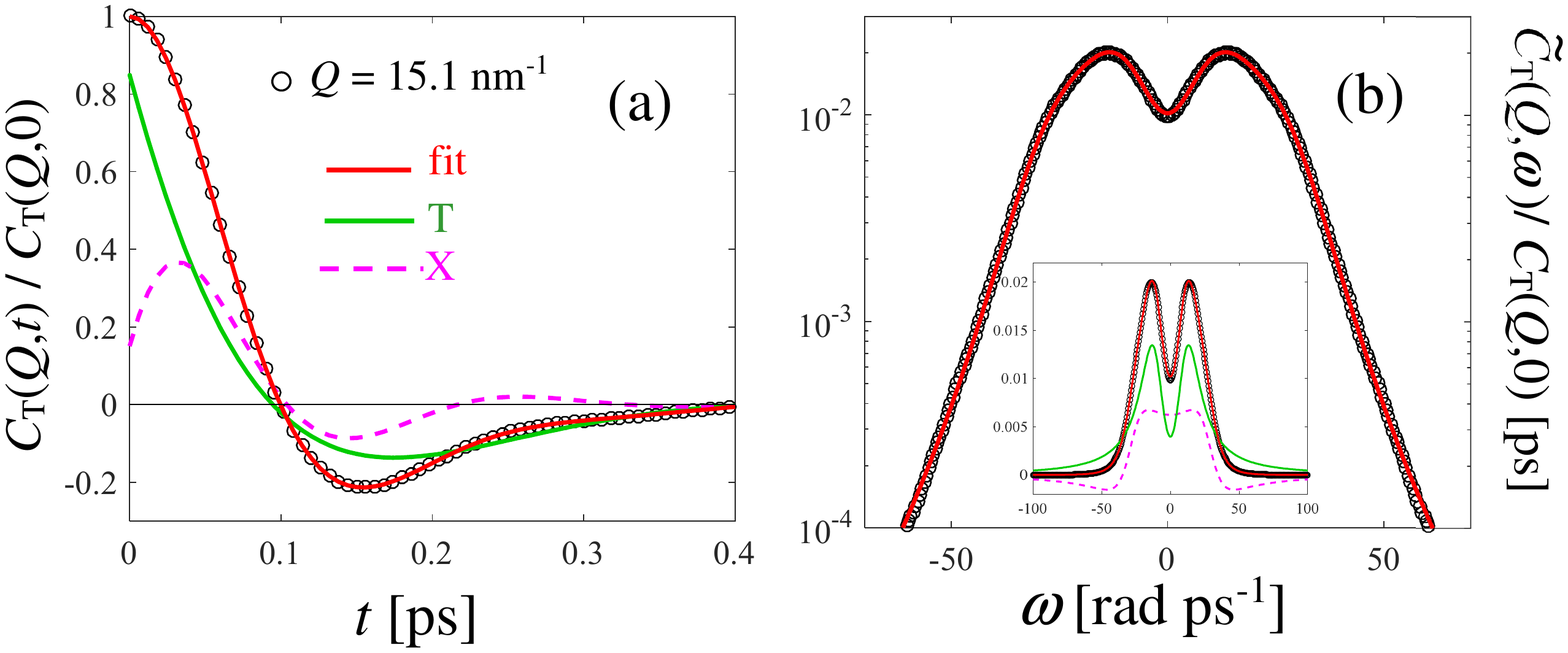}}
\caption{(a) Normalized TCAF of liquid Au at $Q$=15.1 nm$^{-1}$ (black circles) and fit result (red solid curve). The fit components (two complex pairs) are also shown and specified in the legend. (b) Corresponding spectrum and fit results. The semilogarithmic scale helps appreciating the quality of the fit over more than two decades. The inset shows the spectrum and its components in linear scale.}
\label{Fig3}
\end{figure*}

\begin{figure*}
\resizebox{0.95\textwidth}{!}
{\includegraphics[trim=2cm 13cm 2cm 0cm]{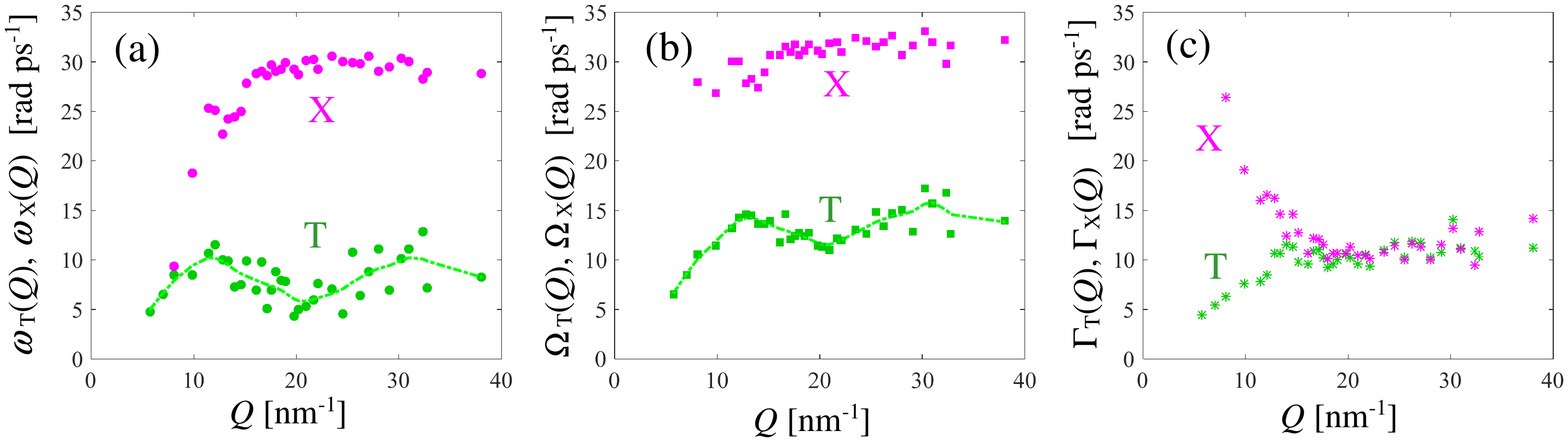}}
\caption{(a) $Q$ dependence of the frequencies $\omega_{\rm T}$ (green full circles) and $\omega_{\rm X}$ (magenta full circles) as obtained from the fits to $C_{\rm T}(Q,t)$. The green dot-dashed spline curve through the T points is just a guide to the eye. (b) Same as panel (a) but for the undamped frequencies $\Omega_{\rm T}$ (green full squares) and $\Omega_{\rm X}$ (magenta full squares). (c) Dampings $\Gamma_{\rm T}$ (green asterisks) and $\Gamma_{\rm X}$ (magenta asterisks). }
\label{Fig4}
\end{figure*}

\begin{figure*}
\resizebox{0.95\textwidth}{!}
{\includegraphics[trim=2cm 10cm 2cm 0cm]{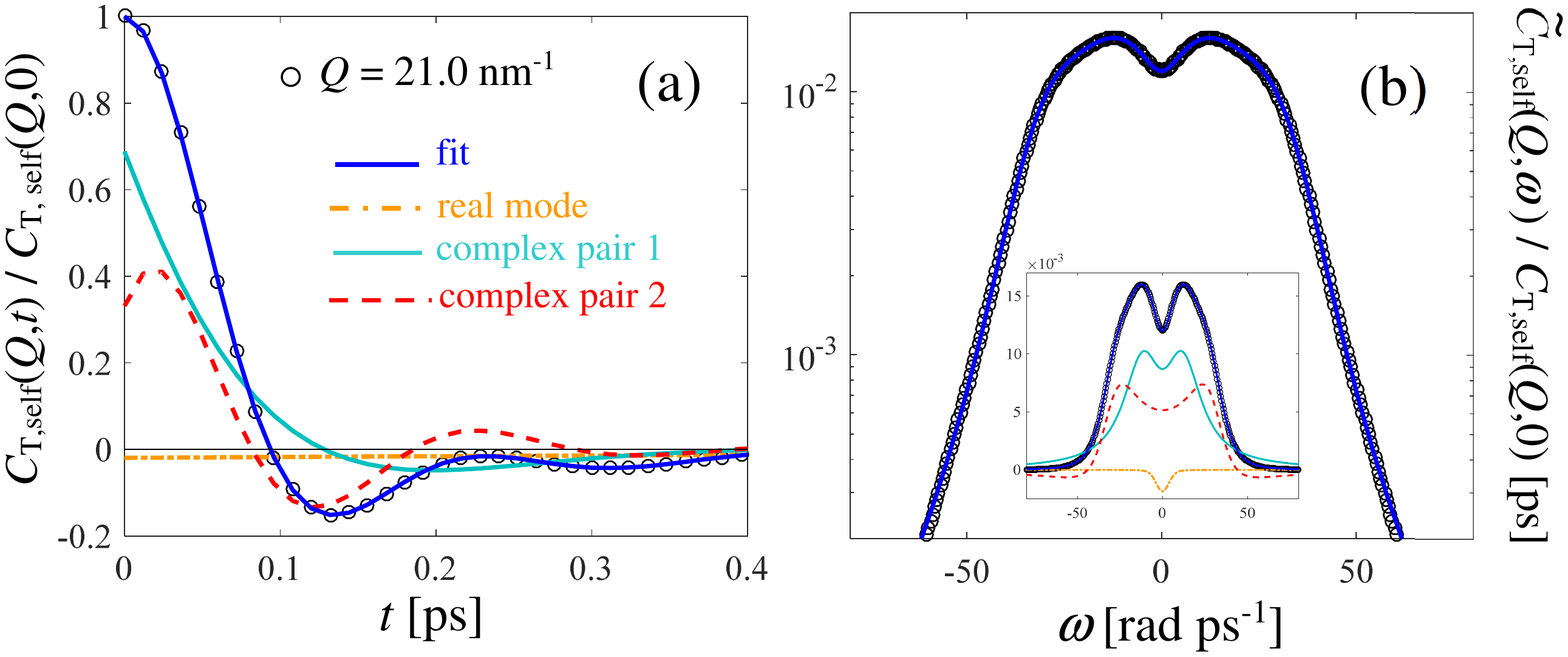}}
\caption{(a) Normalized $C_{\rm T, self}(Q,t)$ of liquid Au at $Q$=21.0 nm$^{-1}$ (black circles) and fit result (blue solid curve). The fit components (two complex pairs plus one real mode) are also shown and specified in the legend. (b) Corresponding spectrum and fit results in semilogarithmic scale. The inset shows the spectrum and all its components in linear scale.}
\label{Fig5}
\end{figure*}

\begin{figure*}
\resizebox{0.95\textwidth}{!}
{\includegraphics[trim=2cm 13cm 2cm 0cm]{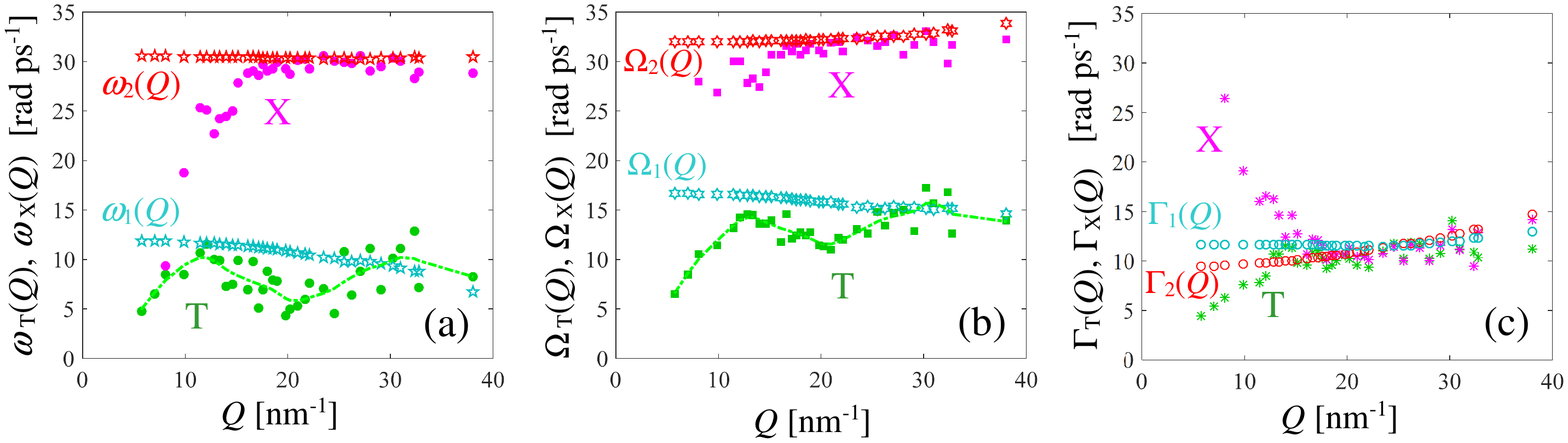}}
\caption{Same as Fig.\ \ref{Fig3} with the addition of the results of the fits to $C_{\rm T,self}(Q,t)$. (a) Empty stars are used for the effective frequencies of the complex pairs labeled as 1 (cyan) and 2 (red) in Fig.\ \ref{Fig4}. (b) Same as panel (a) with equal color code. Empty hexagrams represent $\Omega_1$ and $\Omega_2$. (c) Same as panel (b) but for the dampings. Empty circles are used for $\Gamma_1$ and $\Gamma_2$. }
\label{Fig6}
\end{figure*}

\begin{figure}
\resizebox{0.95\textwidth}{!}
{\includegraphics[trim=2cm 10cm 8cm 1cm]{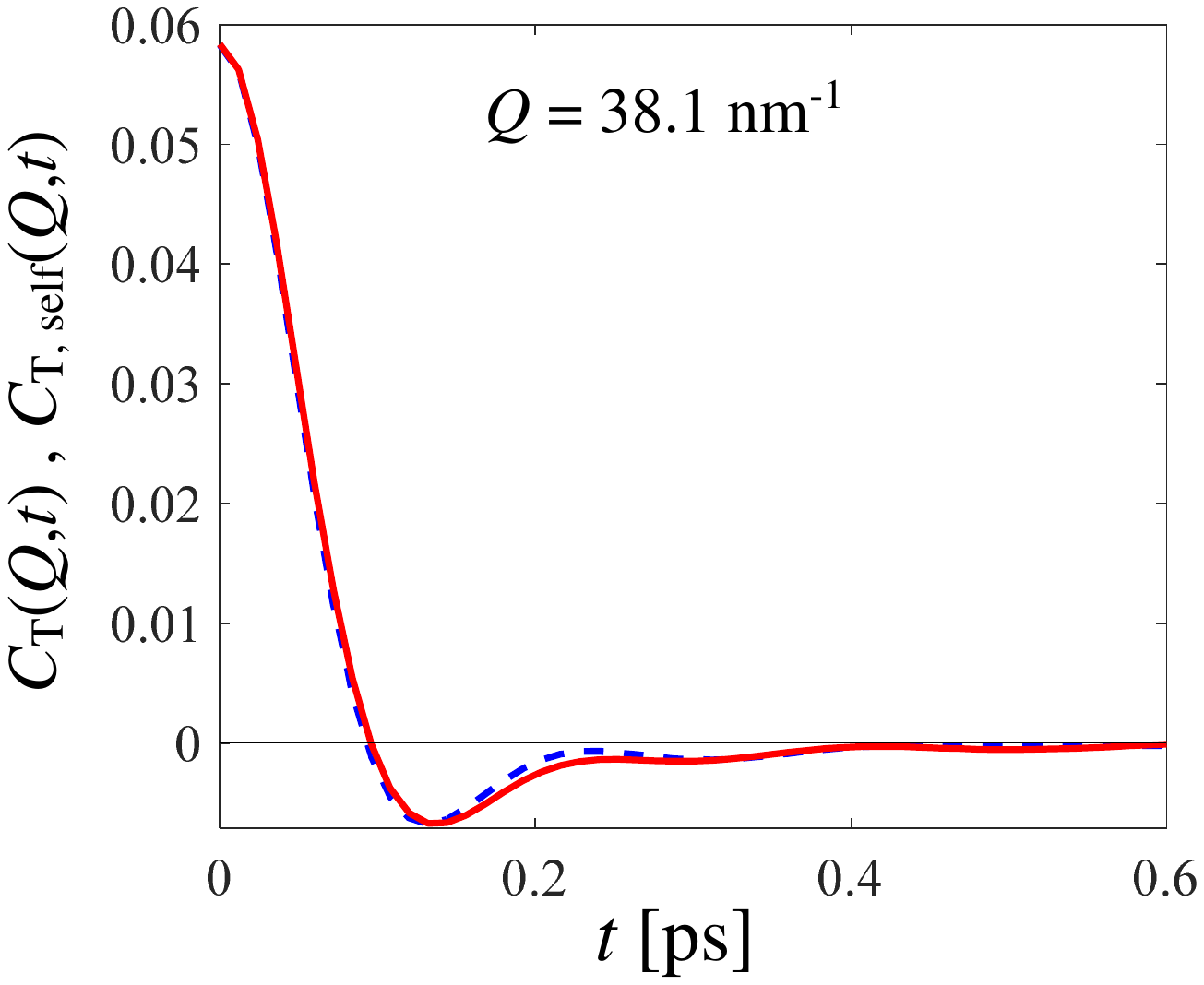}}
\caption{Total (red solid curve) and self (blue dashed curve) TCAF at a high $Q$ value as 38.1 nm$^{-1}$}
\label{Fig7}
\end{figure}

\begin{figure}
\resizebox{0.95\textwidth}{!}
{\includegraphics[trim=2cm 10cm 8cm 1cm]{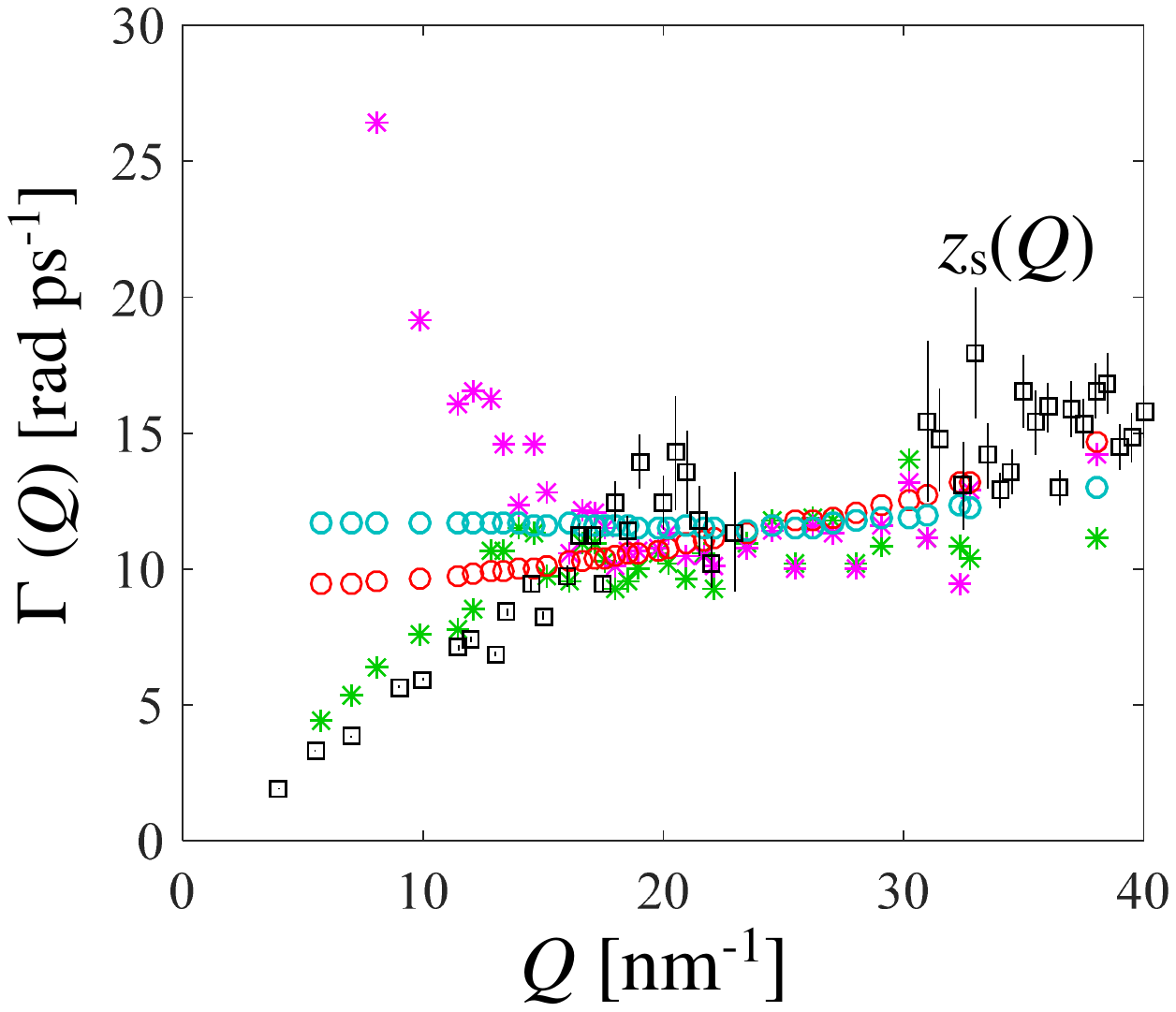}}
\caption{Same as Fig.\ \ref{Fig6}(c) with the addition of the damping $z_{\rm s}(Q)$ (black empty squares with errorbars) of longitudinal excitations as determined from the dynamic structure factor of liquid Au \cite{guarini2013}. Data are missing in the region around $Q_{\rm p}$ because in the propagation gap, where $\omega_{\rm s}(Q) \to 0$, fits to $S(Q,\omega)$ become unstable.}
\label{Fig8}
\end{figure}

Since the VAF is a single-particle (self) property, and the X modes of $C_{\rm T}(Q,t)$ had the same characteristics of the longitudinal contribution to the VAF, we found the calculation and EET analysis of the self part of the TCAF as mandatory, given the fact that $C_{\rm T,self}(Q,t)$ is still quite similar in shape to the VAF at rather high $Q$ values, as shown in Fig.\ \ref{Fig1}. More precisely, although as $Q$ grows $C_{\rm T,self}(Q,t)$ cannot, of course, coincide with the VAF on an absolute scale, it anyway displays the presence of longitudinal waves in a fluid in essentially the same way the VAF does, and in particular with equal frequency, i.e. highlighting a strong similarity in lineshape. We believe this is not a fortuitous coincidence.

The analysis of $C_{\rm T,self}(Q,t)$ revealed to be fundamental, beyond showing once again the effectiveness of the EET in the description of any correlation function, now including also $C_{\rm T,self}(Q,t)$. The importance lies in having shown that the X component, initially of unknown origin, indeed is linked to the single-particle dynamics.

Several other comments, given at the end of the previous section, lead to a few conclusive notes.

The TCAF contains two completely distinguishable signatures of the main propagating waves present in a fluid: one gives evidence of the collective transverse excitation (with its $Q$ dispersion) that the function is appropriate to disclose; the other is, substantially, the ghost of the DoS, unveiling the other (longitudinal) waves in the fluid in the same way the VAF does, i.e., without revealing their true dispersion $\omega_{\rm s}(Q)$, except through the more or less marked damping the VAF shows for a specific excitation. In this sense, the behavior of the X mode of $C_{\rm T}(Q,t)$ at intermediate and high $Q$ values shows that such a component is not a ``new'' and unknown ``propagating collective excitation'' emerging in the TCAF, but is simply, through $C_{\rm T,self}(Q,t)$, the mirror of the longitudinal contribution to the VAF. 

The above observations suggest that what we found for $C_{\rm T}(Q,t)$ should also hold in the reverse case, where studies of $S(Q,\omega)$ (or equivalently $\tilde{C}_{\rm L}(Q, \omega)$ \cite{nota2}) show some ``transverse like'' contribution to it \cite{marques2015,delrio2016,delrio2017,delrio2017a, delriozinco, guarini2020}. The possibility that such an additional transverse signal to $S(Q,\omega)$ might be due to the self part was proposed some years ago in the interpretation of simulation data on liquid Na \cite{garberoglio2018}. As a proof of these guesses, it would be worth analyzing, quantitatively, i.e., by the EET, the transverse signal and its possible relation with the self part also in the case of $S(Q,\omega)$. However, analyses like the present one are very demanding, and cannot be pursued and described in the same paper. Moreover, given the weakly dispersive character of transverse modes, such an analysis would be far less stringent than the present, illuminating, one. 

In conclusion, for both simple monatomic liquids LJ or Au, our picture excludes, as a fact, any real mixing or coupling of the transverse and longitudinal collective excitations in the (different) sense mentioned in many works, rendering this assumption dispensable. More simply, as this study shows, the self dynamics emerges in an evident way, bringing to light also those processes that the longitudinal or transverse character of the studied function should, in principle, forbid to observe.

\begin{acknowledgments}
We thank Emmanuel Farhi for providing the atomic configurations of the AIMD simulation of liquid Au. 
\end{acknowledgments}

\newpage

\newpage

\begin{thebibliography}{}

\bibitem{march} N. H. March, {\it Liquid Metals} (Cambridge University Press, Cambridge, 1990).

\bibitem{balucani} U. Balucani and M. Zoppi, {\it Dynamics of the Liquid State} (Clarendon, Oxford, 1994).

\bibitem{montfrooij} W. Montfrooij and I. de Schepper, {\it Excitations in Simple Liquids, Liquid Metals and Superfluids} (Oxford University Press, New York, 2010).

\bibitem{scopigno_review} T. Scopigno. and G. Ruocco, Microscopic dynamics in liquid metals: The experimental point of view, Rev. Mod. Phys. {\bf 77}, 881 (2005).


\bibitem{guarini2013} E. Guarini, U. Bafile, F. Barocchi, A. De Francesco, E. Farhi, F. Formisano, A. Laloni, A. Orecchini, A. Polidori, M. Puglini, and F. Sacchetti, Dynamics of liquid Au from neutron Brillouin scattering and ab initio simulations: Analogies in the behavior of metallic and insulating liquids, Phys. Rev. B {\bf 88}, 104201 (2013).

\bibitem{marques2015} M. Marqu\'{e}s, L. E. Gonz\'{a}lez and D. J. Gonz\'{a}lez, {\it Ab initio} study of the structure and dynamics of bulk liquid Fe, Phys. Rev. B {\bf 92}, 134203 (2015).

\bibitem{delrio2016} B. G. del Rio, D. J. Gonz\'{a}lez, L. E. Gonz\'{a}lez, An ab initio study of the structure and atomic transport in bulk liquid Ag and its liquid-vapor interface, Phys. Fluids {\bf 28}, 107105 (2016).
 
\bibitem{delrio2017} B. G. del Rio, O. Rodriguez, L. E. Gonz\'{a}lez and D. J. Gonz\'{a}lez, First principles determination of static, dynamic and electronic properties of liquid Ti near melting, Comput. Mater. Sci. {\bf 139}, 243 (2017).

\bibitem{delrio2017a} B. G. del Rio, L. E. Gonz\'{a}lez and D. J. Gonz\'{a}lez, {\it Ab initio} study of several static and dynamic properties of bulk liquid Ni near melting, J. Chem. Phys. {\bf 146}, 034501 (2017).

\bibitem{delriozinco} B. G. del Rio and L. E. Gonz\'{a}lez, Longitudinal, transverse, and single-particle dynamics in liquid Zn:
{\it Ab initio} study and theoretical analysis, Phys. Rev. B {\bf 95}, 224201 (2017).

\bibitem{bellissima2017} S. Bellissima, M. Neumann, E. Guarini, U. Bafile and F. Barocchi, Density of states and dynamical crossover in a dense fluid revealed by exponential mode analysis of the velocity autocorrelation function, Phys. Rev E {\bf 95}, 012108 (2017).

\bibitem{guarini2017}  E. Guarini, S. Bellissima, U. Bafile, E. Farhi, A. De Francesco, F. Formisano and F. Barocchi, Density of states from mode expansion of the self-dynamic structure factor of a liquid metal, Phys. Rev. E {\bf 95}, 012141 (2017).

\bibitem{guarini2020} E. Guarini, A. De Francesco, U. Bafile, A. Laloni, B. G. del Rio,  D. J. Gonz\'{a}lez,  L. E. Gonz\'{a}lez, F. Barocchi and F. Formisano, Neutron Brillouin scattering and ab initio simulation study of the collective dynamics of liquid silver, Phys. Rev. B {\bf 102}, 054210 (2020).

\bibitem{barocchi2012} F. Barocchi, U. Bafile and M. Sampoli, Exact exponential function solution of the generalized Langevin equation for autocorrelation functions of many-body systems, Phys. Rev. E {\bf 85}, 022102 (2012).

\bibitem{barocchi2013} F. Barocchi and U. Bafile, Expansion in Lorentzian functions of spectra of quantum autocorrelations, Phys. Rev. E {\bf 87}, 062133 (2013).

\bibitem{barocchi2014} F. Barocchi, E. Guarini and U. Bafile, Exponential series expansion for correlation functions of many-body systems, Phys. Rev. E {\bf 90}, 032106 (2014).

\bibitem{guarini2023} E. Guarini, M. Neumann, A. De Francesco, F. Formisano, A. Cunsolo, W. Montfrooij, D. Colognesi and U. Bafile, Onset of collective excitations in the transverse dynamics of simple fluids, Phys. Rev. E {\bf 107}, 014139 (2023).

\bibitem{brazhkin} N. P. Kryuchkov, V. V. Brazhkin, and S. O. Yurchenko, Anticrossing of longitudinal and transverse modes in simple fluids, J. Phys. Chem. Lett. {\bf 10}, 4470 (2019).

\bibitem{sampoli97} M. Sampoli, G. Ruocco, and F. Sette, Mixing of longitudinal and transverse dynamics in liquid water, Phys. Rev. Lett. {\bf 79}, 1678 (1997).


%\bibitem{cunsolo} A. Cunsolo, {\it The THz dynamics of liquids probed by inelastic x-ray scattering} (World Scientific, Singapore, 2021). 

\bibitem{guarini2021} E. Guarini {\it et al.}, Collective dynamics of liquid deuterium: Neutron scattering and approximate quantum simulation methods, Phys. Rev. B {\bf 104}, 174204 (2021).

%\bibitem{blairs} S. Blairs, Correlation between surface tension, density, and sound velocity of liquid metals, J. Colloid Interf. Sci. {\bf 302}, 312 (2006).

\bibitem{nota} Regretfully, we have not an estimate of the errors of the simulation data and, consequently, of the fitted parameters. However, errors can be evaluated, in an approximate way, by looking at the spread of the data points.

\bibitem{nota2} It is well known that $\tilde{C}_{\rm L}(Q,\omega)$ is directly related to $S(Q,\omega)$ (see Eq. (1.148) of Ref.\ \cite{balucani}) and that an EET description of the two functions provides the same real and complex modes, differing only in their amplitude (see Appendix A of Ref.\ \cite{guarini2020}) . 

\bibitem{garberoglio2018} G. Garberoglio, R. Vallauri and U. Bafile, Time correlation functions of simple liquids: A new insight
on the underlying dynamical processes, J. Chem. Phys. {\bf 148}, 174501 (2018).

\end{thebibliography}
\end{document}